\documentclass{emulateapj}

\shorttitle{Obscured Star Formation in SLSN Host Galaxies}
\shortauthors{Hatsukade et al.}

\begin{document}

\title{Obscured Star Formation in the Host Galaxies of Superluminous Supernovae}

\author{B.~Hatsukade}
\affiliation{Institute of Astronomy, The University of Tokyo, 2-21-1 Osawa, Mitaka, Tokyo 181-0015, Japan}
\email{hatsukade@ioa.s.u-tokyo.ac.jp}

\author{N.~Tominaga}
\affiliation{Department of Physics, Faculty of Science and Engineering, Konan University, 8-9-1 Okamoto, Kobe, Hyogo 658-8501, Japan}
\affiliation{Kavli Institute for the Physics and Mathematics of the Universe (WPI), The University of Tokyo, 5-1-5 Kashiwanoha, Kashiwa, Chiba 277-8583, Japan}

\author{M.~Hayashi}
\affiliation{National Astronomical Observatory of Japan, 2-21-1 Osawa, Mitaka, Tokyo 181-8588, Japan}

\author{M.~Konishi}
\affiliation{Institute of Astronomy, The University of Tokyo, 2-21-1 Osawa, Mitaka, Tokyo 181-0015, Japan}

\author{Y.~Matsuda}
\affiliation{National Astronomical Observatory of Japan, 2-21-1 Osawa, Mitaka, Tokyo 181-8588, Japan}
\affiliation{Graduate University for Advanced Studies (SOKENDAI), Osawa 2-21-1, Mitaka, Tokyo 181-8588, Japan}

\author{T.~Morokuma}
\affiliation{Institute of Astronomy, The University of Tokyo, 2-21-1 Osawa, Mitaka, Tokyo 181-0015, Japan}

\author{K.~Morokuma-Matsui}
\affiliation{Institute of Space and Astronautical Science, Japan Aerospace Exploration Agency, Chuo-ku, Sagamihara 252- 5210, Japan}

\author{K.~Motogi}
\affiliation{Graduate School of Science and Engineering, Yamaguchi University, Yoshida 1677-1, Yamaguchi, Yamaguchi 753-8512, Japan}

\author{K.~Niinuma}
\affiliation{Graduate School of Science and Engineering, Yamaguchi University, Yoshida 1677-1, Yamaguchi, Yamaguchi 753-8512, Japan}

\author{Y.~Tamura}
\affiliation{Department of Physics, Nagoya University, Furo-cho, Chikusa-ku, Nagoya 464-8602, Japan}


\begin{abstract}
We present the results of 3 GHz radio continuum observations of the 8 host galaxies of super-luminous supernovae (SLSNe) at $0.1 < z < 0.3$ by using the Karl G. Jansky Very Large Array. 
Four host galaxies are detected significantly, and two of them are found to have high star-formation rates (SFRs $>$ 20~$M_{\odot}$~yr$^{-1}$) derived from radio emission, making them the most intensely star-forming host galaxies among SLSN host galaxies. 
We compare radio SFRs and optical SFRs, and find that three host galaxies have an excess in radio SFRs by a factor of $>$2, suggesting the existence of dust-obscured star formation, which cannot be traced by optical studies. 
Two of the three host galaxies, which are located in the galaxy main sequence based on optical SFRs, are found to be above the main sequence based on their radio SFRs. 
This suggests a higher fraction of starburst galaxies in SLSN hosts than estimated in previous studies. 
We calculate extinction from the ratio between radio SFRs and dust-uncorrected optical SFRs and find that the hosts are on the trend of increasing extinction with metallicity, which is consistent with the relation in local star-forming galaxies. 
We also place a constraint on a pulsar-driven SN model, which predicts quasi-steady synchrotron radio emission. 
\end{abstract}


\section{Introduction} \label{sec:introduction}

Superluminous supernovae (SLSNe) are extremely luminous explosions with peak absolute magnitudes of $\lesssim$$-21$~mag, which are $\sim$10--100 times brighter than ordinary Type Ia and core-collapse SNe \citep[][for a review]{gal12}. 
They are detected at high redshifts \citep[$z \sim 4$;][]{cook12}, and therefore can be powerful indicators of environments in the distant universe. 
SLSNe are classified into two main subclasses depending on the presence of hydrogen signatures in the observed spectra: 
hydrogen-poor Type I (SLSN-I) and hydrogen-rich Type II (SLSN-II) \citep[Type I-R is hydrogen-poor events whose light curves are consistent with radioactive decay;][]{gal12}. 
The physical nature of the progenitor of SLSNe is still a matter of debate, especially for SLSNe-I. 
SLSNe-II are likely to be explained by a shock between the SN ejecta and surrounding dense hydrogen-rich circumstellar medium \citep[e.g.,][]{woos07, mori13}. 
On the other hand, several progenitor and explosion models have been proposed for SLSNe-I 
such as 
pair-instability SN \citep[e.g.,][]{gal09}, 
SN which produces a large amount of $^{56}$Ni \citep[e.g.,][]{mori10}, 
spin-down of a newborn strongly magnetic neutron star \citep[magnetar; e.g.,][]{kase10, woos10}, 
fallback accretion onto a compact remnant \citep{dext13}, 
and interaction with dense circumstellar medium \citep[e.g.,][]{chev11, soro16}.

In order to constrain the progenitor models, it is essential to understand the properties of their host galaxies. 
Previous studies have shown that SLSN-I hosts are typically dwarf galaxies with low-luminosity, low stellar mass, low star-formation rate (SFR), and high specific SFR (sSFR) compared to local star-forming galaxies and the hosts of core-collapse SNe, while SLSN-II hosts show a wider range than SLSN-I hosts \citep[e.g.,][]{lunn14, lelo15, angu16, perl16}. 
SLSN hosts have also been compared to those of long-duration gamma-ray bursts (LGRBs), which are thought to be originated from the explosion of massive stars \citep[e.g.][]{hjor03, stan03}. 
While \cite{lunn14} and \cite{jape16} suggest that SLSNe-I and LGRB hosts are similar in terms of SFR, stellar mass, and sSFR, \cite{lelo15} and \cite{angu16} argue that SLSNe-I hosts have lower stellar mass and SFR.

To understand the environment forming SLSN progenitors, the accurate estimate of star-forming activity is essential. 
An important factor to be considered is the effect of obscuration by dust. 
The observations of SLSN hosts have been made exclusively in the optical/near-infrared (NIR) wavelengths, which are subject to dust extinction in contrast to longer wavelengths, and it is possible that we are missing dust-obscured star formation in SLSN hosts. 
In this respect radio observations are important to probe dust-obscured star formation. 
Recently, \cite{schu18} searched radio emission for a sample of SLSN hosts from the survey data of Faint Images of the Radio Sky at Twenty-Centimeters \citep[FIRST;][]{beck95} with the Karl G. Jansky Very Large Array (VLA), the NRAO VLA Sky Survey \citep[NVSS;][]{cond98}, and the Sydney University Molonglo Sky Survey \citep[SUMSS;][]{bock99}.  
No host is detected in the surveys down to the rms levels of $\sim$0.15, $\sim$0.45, and $\sim$1.3 mJy~beam$^{-1}$, respectively. 
They also conducted deeper VLA observations of three hosts of SLSNe at $z = 0.1$--0.3 (MLS121104, SN~2005ap, and SN~2008fz), and obtained upper limits with the rms noise level of 15, 25, and 15 $\mu$Jy~beam$^{-1}$, respectively. 
The number of SLSN hosts with deep radio observations is still very limited, and it is essential to study a larger sample.

Radio observations are also useful to constrain progenitor models which predict quiescent radio emission from SLSN remnants on timescales of decades \citep{mura16, kash17, metz17}. 
Based on the pulsar-driven SN model of \cite{mura16}, \cite{oman18} predict quasi-state synchrotron radio emission peaking at $\gtrsim$$10$ years after the SN explosion with the expected radio emission of $>$5--10~$\mu$Jy at 1~GHz for some of the known brightest SLSNe-I, which can dominate radio emission from the hosts and is tested with current radio telescopes.

In this paper, we present the results of 3-GHz radio continuum observations of 8 SLSN hosts by using the VLA. 
Section~\ref{sec:observations} describes target SLSN hosts, VLA observations, and data reduction. 
The results are shown in Section~\ref{sec:results}. 
In Section~\ref{sec:discussion}, we discuss obscured star formation in the hosts and constraint on progenitor models. 
Conclusions are presented in Section~\ref{sec:conclusions}. 
Throughout the paper, we adopt cosmological parameters of $H_0=67.8$ km s$^{-1}$ Mpc$^{-1}$, $\Omega_{\rm{M}}=0.308$, and $\Omega_{\Lambda}=0.692$ based on the results of full-mission {\sl Planck} observations \citep{plan16}. 
SFRs and stellar masses are converted to a \cite{chab03} IMF from a \cite{salp55} IMF or a \cite{krou01} IMF by multiplying a factor of 0.61 and 1.1, respectively \citep[e.g.,][]{mada14}.

\begin{table*}
\begin{center}
\caption{Properties of the Targets}
\label{tab:targets}
\begin{tabular}{ccccccccc}
\hline
SLSN & Class &$z$&R.A.$^{\rm a}$&Decl.$^{\rm a}$&SFR(SED)$^{\rm b}$ &SFR(H$\alpha)^{\rm c}$ & $\log(M_*)^{\rm d}$ & $12+\log({\rm O/H})^{\rm e}$ \\
     &       &   &(J2000)       &(J2000)        &($M_{\odot}$~yr$^{-1}$)&($M_{\odot}$~yr$^{-1}$)& ($M_{\odot}$)      & \\
\hline                                                                                                                 
PTF 12dam &I-R&0.107&14:24:46.20&$+$46:13:48.3&$11.13^{+3.376}_{-3.339}$&$4.781^{+0.965}_{-1.174}$&$ 8.30^{+0.15}_{-0.15}$&$8.00^{+0.01}_{-0.01}$ \\
SN 1999bd &II &0.151&09:30:29.17&$+$16:26:07.8&                         &$1.09\pm0.34$            &$ 9.52^{+0.26}_{-0.24}$&$8.52 \pm 0.02$ \\
PTF 11rks &I  &0.192&01:39:45.53&$+$29:55:27.4&$1.064^{+0.346}_{-0.429}$&$0.389^{+0.202}_{-0.147}$&$ 9.11^{+0.13}_{-0.16}$&$8.17^{+0.11}_{-0.17}$ \\
PTF 10aagc&I  &0.206&09:39:56.92&$+$21:43:17.1&$1.566^{+1.049}_{-0.646}$&$0.474^{+0.187}_{-0.160}$&$ 8.98^{+0.13}_{-0.21}$&$8.19^{+0.04}_{-0.05}$ \\
SN 2010gx &I  &0.230&11:25:46.71&$-$08:49:41.4&$0.532^{+0.287}_{-0.248}$&$0.257^{+0.052}_{-0.051}$&$ 7.87^{+0.13}_{-0.21}$&$7.94^{+0.09}_{-0.14}$ \\
SN 2008am &II &0.234&12:28:36.30&$+$15:34:50.0&                         &$1.38\pm0.39$            &$ 9.13^{+0.19}_{-0.14}$&$8.35 \pm 0.02$ \\
PTF 10qaf &II &0.284&23:35:42.89&$+$10:46:32.9&                         &$3.13\pm0.89$            &$10.24^{+0.22}_{-0.17}$&$8.68 \pm 0.04$ \\
PTF 10uhf &I  &0.288&16:52:46.70&$+$47:36:21.8&$6.837^{+2.227}_{-3.103}$&$19.36^{+7.301}_{-5.764}$&$11.23^{+0.12}_{-0.15}$&$8.70^{+0.01}_{-0.01}$ \\
\hline 
\multicolumn{9}{@{}l@{}}{\hbox to 0pt{\parbox{170mm}
{
\smallskip
$^{\rm a}$ SLSN position used as a phase center of the VLA observations. \\
$^{\rm b}$ Extinction-corrected SFR calculated from UV--optical--NIR SED fit. \\
$^{\rm c}$ Extinction-corrected SFR calculated from H$\alpha$ flux. \\
$^{\rm d}$ Stellar mass calculated from UV--optical--NIR SED fit. \\
$^{\rm e}$ Metallicity based on the \cite{pett04} O3N2 diagnostic. \\
References for SFR, stellar mass, and metallicity are \citet{perl16} for PTF 12dam, PTF 11rks, PTF 10aagc, SN 2010gx, and PTF 10uhf, and \citet{lelo15} for SN 1999bd, SN 2008am, and PTF 10qaf. 
}\hss}}
\end{tabular}
\end{center}
\end{table*}

\section{VLA Observations} \label{sec:observations}

\subsection{Targets} \label{sec:targets}

We concentrate on SLSN hosts which have 
(i) an accurately determined redshift, 
and (ii) a measured optical SFR. 
Targets are selected from comprehensive studies of SLSN hosts in literature \citep{lunn14, lelo15, angu16, perl16}. 
From the sample of SLSN hosts, we selected the hosts with SFR $>$ 1~$M_{\odot}$~yr$^{-1}$
\citep[and also include the SN 2010gx host which is reported to have one of the highest SFR among the sample of][]{angu16} 
 and at sufficiently low redshifts ($z < 0.3$) to ensure significant constraint on obscured star formation. 
In oder to avoid the contamination from AGN to radio emission, we exclude SLSN hosts which are known to have possible AGN features. 
We also confirmed that the targets are not listed in the X-ray source catalog of NASA's High Energy Astrophysics Science Archive Research Center (HEASARC)\footnote{https://heasarc.gsfc.nasa.gov/docs/software.html}. 
The properties of the targets are presented in Table~\ref{tab:targets}.

While \cite{perl16} show that PTF~10qaf occurred in a companion galaxy (SFR $= 0.268$~$M_{\odot}$~yr$^{-1}$) $\sim$$4''$ away from a nearby spiral galaxy at the same redshift, the galaxy pair is studied by \cite{lelo15} as a single object with SFR $= 3.13$~$M_{\odot}$~yr$^{-1}$.
The VLA beamsize of our observations ($8\farcs1 \times 6\farcs1$) does not resolve the galaxy pair and we treat them as a single system.

Some of the hosts have higher SFRs derived from SED fit than those derived from H$\alpha$ flux, suggesting that the Balmer decrement underestimates the extinction compared to that estimated from the SED fit. 
Hereafter we use SFR(SED) of \citet{perl16} if available, otherwise we use SFR(H$\alpha$) of \citet{lelo15} presented in Table~\ref{tab:targets}. 
We note that the derivation methods of SFRs could have systematic uncertainties. 
The method of SED fit has systematic uncertainties due to assumptions such as extinction, star formation history, and initial mass function \citep{perl13}.

\subsection{Observations and Data Reduction} \label{sec:reduction}

The VLA S-band 3-GHz (13-cm) observations (Project ID: 17A-140) were performed on May 28 and 29, 2017 (5--8 years after the maximum date of the SLSNe) using 27 antennas in the C array configuration. 
The baseline length ranges from 44.8~m to 3.4~km. 
The WIDAR correlator was used with 8-bit samplers. 
We use two basebands with 1 GHz bandwidth centered at 2.5 GHz and 3.5 GHz, which provides a total bandwidth of 2 GHz. 
The field-of-view is $7\farcm4$ (full width at half power). 
The positions of the SNe are used as phase centers. 
Bandpass and amplitude calibrations were done with 3C286 or 3C48, and phase calibrations were done with nearby quasars. 
The total observing time of each target is 1.5 hours.

The data were reduced with Common Astronomy Software Applications \citep[CASA;][]{mcmu07} release 4.7.2. 
About 10\%--25\% of the data were flagged by the pipeline processing. 
The maps were produced with the task {\verb tclean } down to 
a threshold of twice the rms noise level measured in a source-free region in the dirty map 
with the parameters {\verb cell } of 1 arcsec, {\verb gridder } of standard, {\verb specmode } of multi-frequency synthesis, and {\verb nterms } of 2. 
The Briggs weighting with {\verb robust } 0.5 is adopted. 
The resultant synthesized beamsize is $\sim$$6''$--$9''$ (Table~\ref{tab:radio}). 
The absolute flux accuracy is estimated by comparing the measured flux density of the amplitude calibrators and the flux density scale of \cite{pb17}, and the difference is found to be $<$5\%. 
The typical rms noise level of the maps is 4--7~$\mu$Jy~beam$^{-1}$, which is estimated by fitting the pixel--flux histogram of the map with a Gaussian. 
The local rms noise levels around the hosts are estimated with the {\sc BANE} program \citep{hanc12}, which performs $3\sigma$ clipping in the map and calculate the standard deviation on a sparse grid of pixels and then interpolate to make a noise image. 
The local rms noise levels are 5--13~$\mu$Jy~beam$^{-1}$. 
The difference in the rms noise levels measured in the two ways is less than twice the rms value estimated by fitting the pixel--flux histogram, which can be explained by a map fluctuation. 
We adopt the local rms noise levels for deriving physical quantities of the hosts. 
When a source is spatially resolved by the synthesized beam, we measure an integrated flux density by using the {\verb imfit } task, otherwise we adopt a peak intensity.

\begin{figure*}
\begin{center}
\includegraphics[width=.9\linewidth]{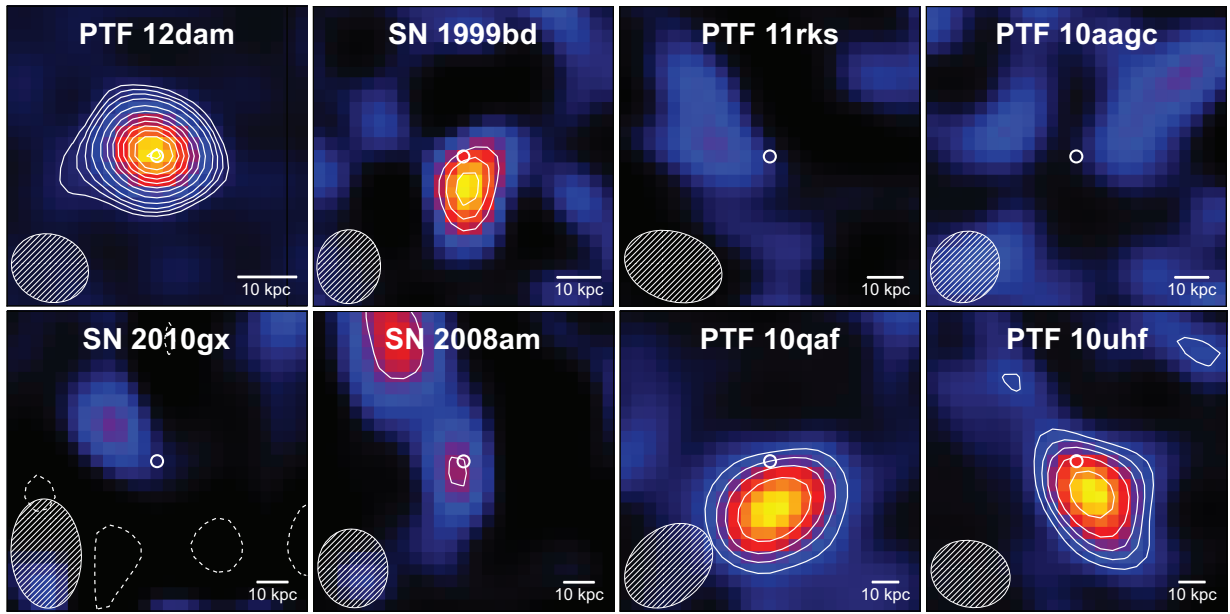}
\end{center}
\caption{
VLA 3-GHz continuum maps centered at the SN position. 
The image size is $25'' \times 25''$. 
North is up and east is to the left. 
White circles represent the SN position.
Contours are $-2.5\sigma$, $2.5\sigma$, $3.5\sigma$, $4.5\sigma$, and $2.5\sigma$ steps subsequently (negative contours as dashed). 
The synthesized beam size is shown in the lower left corners. 
}
\label{fig:radio}
\end{figure*}

\begin{figure*}
\begin{center}
\includegraphics[width=.9\linewidth]{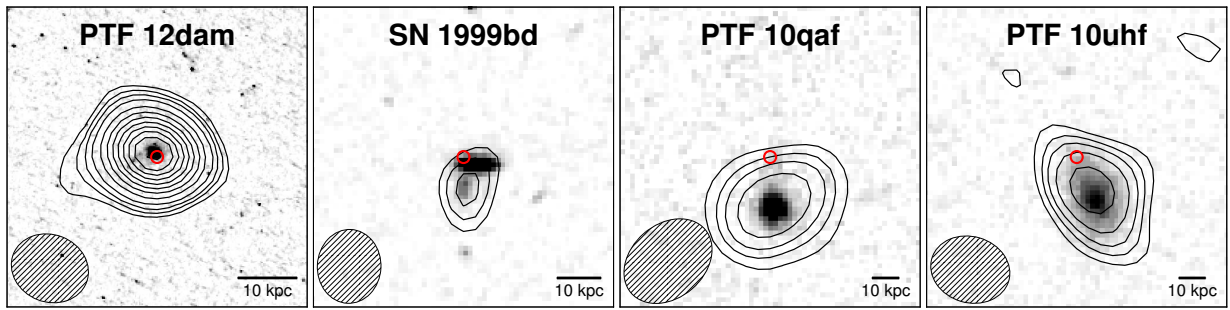}
\end{center}
\caption{
VLA 3-GHz contours overlaid on optical/NIR images for the radio-detected hosts ({\sl HST} WFC3/UVIS F336W for PTF~12dam, WFC3/IR F160W for SN~1999bd, SDSS $r'$ for PTF~10qaf and PTF~10uhf). 
The image size is $25'' \times 25''$. 
Red circles represent the SN position.
Contour levels are the same as in Figure~\ref{fig:radio}. 
}
\label{fig:optical}
\end{figure*}

\section{Results} \label{sec:results}

We detected radio emission in four SLSN hosts (PTF~12dam, SN~1999bd, PTF~10qaf, PTF~10uhf) with a peak signal-to-noise ratio (S/N) above 5.
The radio continuum images are shown in Figure~\ref{fig:radio}. 
The hosts of PTF~10qaf and PTF~10uhf are spatially resolved with the synthesized beam, while the other hosts are only marginally resolved or not resolved.

Figure~\ref{fig:optical} shows the radio contours overlaid on the optical/NIR images for the radio-detected hosts. 
The radio emission in the hosts of SN~1999bd, PTF~10qaf, and PTF~10uhf is dominated by the galaxy center rather than the SN positions a few arcsec away from the center. 
The radio emission in the PTF~10uhf host is elongated toward the northwest of the galaxy. 
This can be caused by a faint companion galaxy $2''$ northwest of the primary nucleus, where \citet{perl16} suggests a merger of a massive spiral galaxy with a less massive disk galaxy. 
While the peak radio position of the SN~1999bd host appears to be associated with a faint object $\sim$$2''$ south of the SN position, we use the whole radio emission for the flux density of the SN~1999bd host system by considering the spatial resolution of the radio observations.

The host of SN~2008am is only tentatively detected (S/N = 2.6) and the hosts of PTF~11rks, PTF~10aagc, and SN~2010gx are not detected. 
We derived 3$\sigma$ upper limits on SFRs for those hosts in the subsequent section.

The radio-detected hosts lie at the location of star-forming galaxies in the BPT diagram \citep{lelo15, perl16} and are not listed in the X-ray catalog, suggesting that the radio emission is primarily powered by star formation. 
Although there could be a Compton-thick AGN, it is difficult to further investigate it with existing data sets. 
In this paper we assume that the radio emission is dominated by star-forming activity.
SFRs based on the radio emission are derived by using the equation of \cite{murp11} for 1.4-GHz flux densities, which is used in previous studies of LGRB host galaxies \citep[e.g.,][]{pp13, perl15, grei16}. 
\cite{grei16} extrapolate flux densities from the rest-frame frequency to 1.4 GHz and provide the equation of radio-derived SFRs as
\begin{eqnarray}
{\rm SFR} = 0.059 S_{\nu} (1+z)^{-(\alpha+1)} {D_L}^2 \nu^{-\alpha}, 
\end{eqnarray}
where $S_{\nu}$ is the observed flux density in ${\rm \mu Jy}$, $\nu$ is the observing frequency in GHz, $\alpha$ is the synchrotron spectral index, and $D_L$ is the luminosity distance in Gpc. 
The synchrotron spectral index $\alpha$ is known to lie between around $-0.8$ and $-0.7$ \citep[e.g.,][]{gioi82, duri88, cond92, nikl97, taba17}, and we adopt $\alpha = -0.75$ following previous studies of LGRB hosts \citep{pp13, perl15, grei16}. 
The derived SFRs and sSFRs are shown in Table~\ref{tab:radio}. 
Note that SFRs would change by a factor of 1.4 and 0.7 if we assume $\alpha$ of $-1.0$ and $-0.5$, respectively. 
The hosts of PTF~10qaf and PTF~10uhf have high SFRs ($>$$20~M_{\odot}$~yr$^{-1}$), making them the most intensely star-forming hosts among SLSN hosts.

\section{Discussions} \label{sec:discussion}

\subsection{Obscured Star Formation} \label{sec:sfr}

\begin{figure}
\begin{center}
\includegraphics[width=\linewidth]{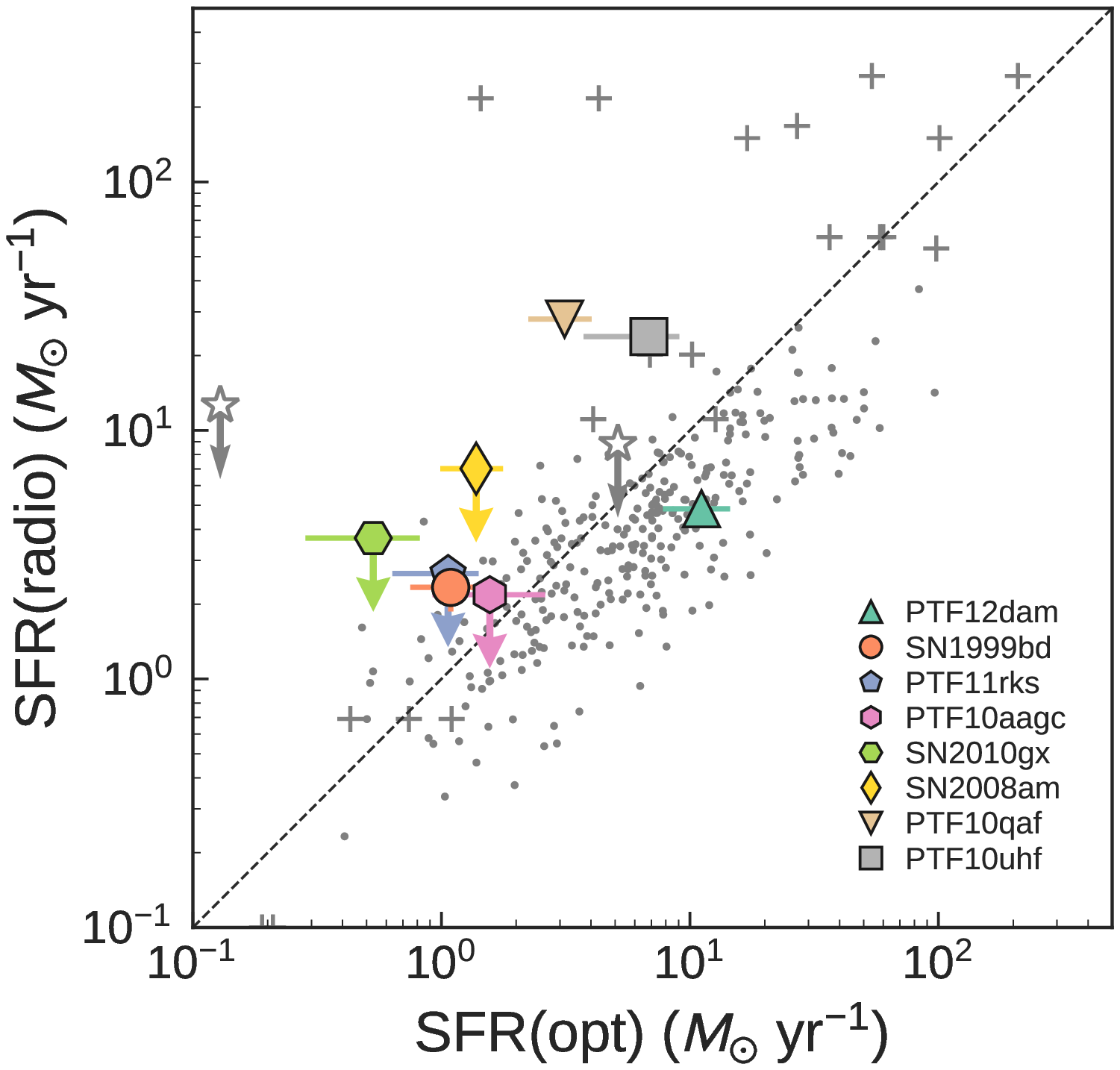}
\end{center}
\caption{
Comparison of SFRs derived from optical and radio observations. 
The dashed line represents the one to one relation. 
Stars represent the SLSN hosts based on radio observations by \cite{schu18}.
We also plot LGRB host galaxies with radio detection (crosses) compiled by \cite{grei16} 
and star-forming galaxies without AGN feature from the VLA-COSMOS survey source catalog \citep{smol17a, smol17b} at $0.1 < z < 0.3$ (dots) as a control sample. 
Arrows represent 3$\sigma$ upper limits. 
}
\label{fig:sfr-sfr}
\end{figure}

We compare the SFRs of the SLSN hosts derived from the radio observations and optical observations (corrected for extinction) in Figure~\ref{fig:sfr-sfr}. 
The hosts of SN~1999bd, PTF~10qaf, and PTF~10uhf have an excess of the radio-derived SFR over optically-derived SFR by a factor of 2--9, suggesting that there exists obscured star formation which cannot be traced by the previous optical/NIR studies. 
The hosts of PTF~10qaf and PTF~10uhf are likely to be interacting \citep{perl16, ciko17}, which could induce dusty star formation. 
The cause of the radio excess for the SN~1999bd host may also be due to an interaction, which appears in the optical image, or a contamination from a nearby source to the radio emission.
Although the sample size is limited, we do not find differences in obscured star formation between the hosts of SLSN-I and SLSN-II. 
We also compare with the hosts of LGRBs with radio observations \citep{stan10, stan14, stan15, hats12, mich12, mich15, pp13, perl15, perl17, grei16}. 
The majority of LGRB hosts are not detected in the radio (not plotted in Figure~\ref{fig:sfr-sfr}) and have less dust-obscured SFR \citep{grei16}, although several LGRB hosts (out of about 60 hosts) have an excess of radio SFR. 
In order to study whether or not SLSN hosts with significant dust-obscured star formation are rare exceptions as in LGRB hosts, a larger sample is needed.

\begin{figure}
\begin{center}
\includegraphics[width=.95\linewidth]{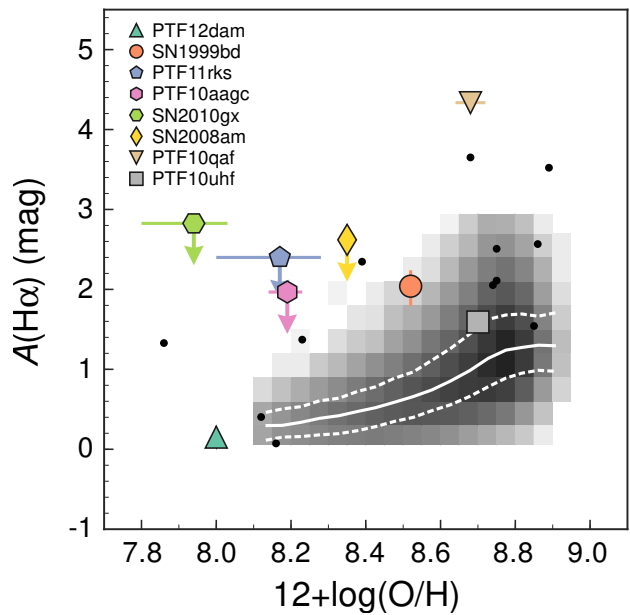} 
\end{center}
\caption{
Extinction at H$\alpha$ wavelength as a function of metallicity. 
Arrows represent 3$\sigma$ upper limits. 
Star-forming galaxies in the sample of \citet{klei17} are plotted as black dots. 
The grey-scale indicates the number of galaxies in the sample of SDSS star-forming galaxies in \cite{garn10}, and the solid and dashed lines indicate the median and standard deviation of the sample, respectively. 
}
\label{fig:av}
\end{figure}

Assuming that the radio emission traces the total star-forming activity in the hosts, we calculate the extinction at H$\alpha$ wavelength from the ratio between the radio SFRs and dust-uncorrected SFRs derived from H$\alpha$ flux as 
\begin{eqnarray}
A({\rm H}\alpha) = -2.5 \log \left( \frac{\rm SFR(radio)}{{\rm SFR(H}\alpha)} \right) {\rm mag}. 
\end{eqnarray}
We plot the extinction as a function of metallicity in Figure~\ref{fig:av}, and find a trend of increasing extinction with increasing metallicity. 
Lower-metallicity galaxies are expected to have a smaller amount of dust, and the correlation has been reported for star-forming galaxies \citep[e.g.,][]{garn10}, and in a recent study by \citet{klei17} based on radio continuum observations. 
While the tendency seen in the SLSN hosts are consistent with star-forming galaxies, the PTF~10qaf host has a higher extinction compared to the sample of Sloan Digital Sky Survey galaxies in \cite{garn10}, suggesting that it has highly obscured star formation. 
This may be due to interaction suggested from optical observations \citep{perl16}.

\begin{figure}
\begin{center}
\includegraphics[width=\linewidth]{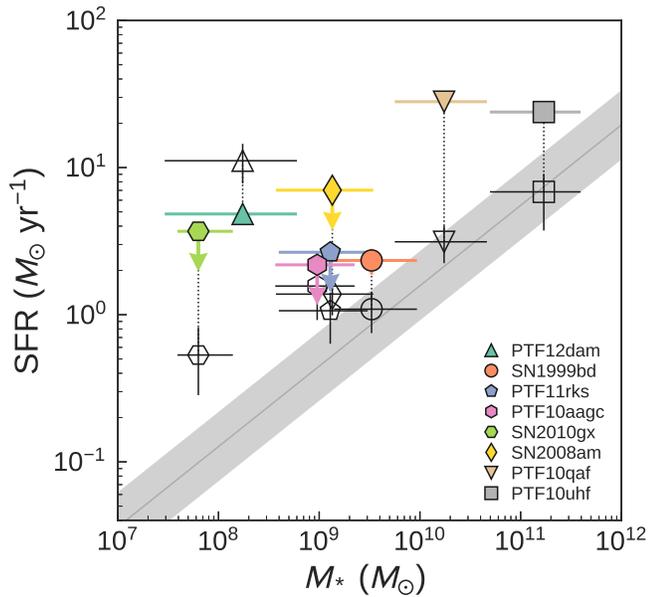}
\end{center}
\caption{
SFRs as a function of stellar mass. 
Open and filled symbols represent the optically-derived SFRs and radio-derived SFRs, respectively. 
Arrows represent 3$\sigma$ upper limits. 
The solid curve and the shaded region show the main-sequence of star-forming galaxies \citep{spea14} at $z = 0.2$ and its uncertainty ($\pm 0.2$~dex). 
}
\label{fig:sfr-ms}
\end{figure}

Figure~\ref{fig:sfr-ms} compares the stellar mass and SFRs for the SLSN hosts. 
It is known that star-forming galaxies follow a tight correlation between stellar mass and SFR, referred to as galaxy main sequence.
Although the hosts of SN~1999bd, PTF~10qaf, and PTF~10uhf are located within the range of the main sequence based on the previous optical observations, the radio observations find that they are above the main sequence, suggesting that they have a starburst nature \citep[e.g.,][]{rodi11, elba11}. 
The hosts of PTF~12dam and PTF~10qaf have sSFR $> 10^{-9}$~yr$^{-1}$, which are higher among SLSN hosts \citep{lelo15, perl16}
\footnote{
Note that the effect of dust attenuation on stellar mass derived from SED fitting is in general small even for dusty galaxies \citep[$<$0.2 dex on average;][]{mich14}, and our result does not change significantly even if we adopt the attenuation estimated from the ratio between SFR(radio) and SFR(H$\alpha$). 
}. 
\cite{perl16} found a higher fraction of starbursts in their SLSN host sample (3--6 SLSN-I hosts out of 18 and 0--2 SLSN II hosts out of 13) with sSFR $> 2 \times 10^{-9}$~yr$^{-1}$ compared to a local comparison sample. 
Our finding of high sSFR hosts based on radio observations supports the higher starburst fraction in SLSN hosts. 

\begin{table*}
\begin{center}
\caption{Results of Radio Observations}
\label{tab:radio}
\begin{tabular}{cccccccc}
\hline
SLSN & Beamsize & Local RMS & S/N$_{\rm peak}$ & Flux Density & SFR(radio) &sSFR$^{\rm a}$ &$A({\rm H}\alpha)^{\rm b}$\\
     & ($''$)   & ($\mu$Jy beam$^{-1}$) & & ($\mu$Jy) & ($M_{\odot}$~yr$^{-1}$) & (yr$^{-1}$) & (mag) \\
\hline
PTF~12dam&$6.5\times5.6$& 5.1&27.7&$141.5\pm5.1$&$4.8\pm0.2$ &$(2.4^{+1.0}_{-0.7})\times10^{-8}$ &$0.15\pm0.04$\\
SN~1999bd&$6.2\times5.3$& 6.5& 5.0&$ 32.8\pm6.5$&$2.3\pm0.5$ &$(7.1^{+6.0}_{-3.3})\times10^{-10}$&$2.0\pm0.2$\\
PTF~11rks&$8.3\times5.8$& 7.3&-   &$<$22.0      &$<$2.7      &$<$$2.1\times 10^{-9}$             &$<$2.4\\
PTF~10aag&$6.1\times5.6$& 5.2&-   &$<$15.6      &$<$2.2      &$<$$2.3\times 10^{-9}$             &$<$2.0\\
SN~2010gx&$9.1\times5.9$& 6.9&-   &$<$20.7      &$<$3.7      &$<$$5.0\times 10^{-8}$             &$<$2.8\\
SN~2008am&$6.6\times5.8$&12.6&-   &$<$37.9      &$<$7.0      &$<$$5.2\times 10^{-9}$             &$<$2.6\\
PTF~10qaf&$8.1\times6.1$& 8.3& 8.5&$ 98.3\pm7.2$&$28.0\pm2.1$&$(1.6^{+1.1}_{-0.5})\times10^{-9}$ &$4.3\pm0.1$\\
PTF~10uhf&$6.7\times5.5$& 5.6&10.5&$ 80.9\pm5.9$&$23.8\pm1.7$&$(1.4^{+0.5}_{-0.4})\times10^{-10}$&$1.6\pm0.1$\\
\hline
\multicolumn{8}{@{}l@{}}{\hbox to 0pt{\parbox{160mm}
{
\smallskip
Limits are 3$\sigma$. \\
$^{\rm a}$ Specific SFR based on SFR(radio). \\
$^{\rm b}$ Extinction at H$\alpha$ wavelength derived from the ratio between SFR(radio) and SFR(H$\alpha)$. 
}\hss}}
\end{tabular}
\end{center}
\end{table*}

\subsection{Constraint on Pulsar-driven SLSN Model} \label{sec:model}

Finally, we discuss the constraint on a theoretical model of SLSNe. 
\cite{mura16} show in their model that pulsar-driven SN remnants cause quasi-steady synchrotron radio emission associated with non-thermal electron-positron pairs in nascent pulsar wind nebulae (PWNe) on a timescale of decades, which may be bright enough to be detected with current radio telescopes. 
Based on the model of \cite{mura16} and \cite{kash17}, \cite{oman18} fit the pulsar-driven SN model to the light curves of six known brightest SLSNe-I including one of our targets, SN~2010gx, and calculated radio emission with the obtained parameters of initial spin period, magnetic field strength, and ejecta mass. 
They calculated radio light curves at 1~GHz and 100~GHz in two cases: with maximum absorption and with no absorption processes in the PWN and SN ejecta. 
The radio emission can be absorbed in the PWN and the SN ejecta, but the system can be transparent at $\gtrsim$10 years. 
The model shows that the radio emission increases with time, reaches its peak at $\sim$10--30 years after the explosion, and then decreases. 
The predicted radio emission of SN~2010gx for the case of no absorption processes is $\sim$60~$\mu$Jy at 3 GHz at the time of our radio observations (seven years after the SN explosion) if we simply assume a synchrotron self-absorption spectral index of 2.5. 
This is higher than our 3$\sigma$ upper limit of $<$21~$\mu$Jy, suggesting that the model with no absorption processes is inconsistent with our observations. 
For the case of maximum absorption, the predicted flux density is well below our detection limit and it is not possible to make a conclusion. 
Follow-up observations and long-term monitoring ($\gtrsim$10 years) are important to constrain the radio light curves and progenitor models with different parameters.

\section{Conclusions}\label{sec:conclusions}

We performed VLA 3-GHz observations of the hosts galaxies of 8 SLSNe ($0.1 < z < 0.3$), 
and detected four SLSN hosts (PTF~12dam, SN~1999bd, PTF~10qaf, PTF~10uhf). 
We derived SFRs from the radio emission, and found that the hosts of PTF~10qaf and PTF~10uhf have high SFRs ($>$$20~M_{\odot}$~yr$^{-1}$), making them the most intensely star-forming galaxies among SLSN hosts. 
Comparison between radio SFRs and optical SFRs shows that the hosts of SN~1999bd, PTF~10qaf, and PTF~10uhf have an excess of radio SFRs over optical SFRs by a factor of $>$2, suggesting that obscured star formation exist in the hosts which cannot be traced by the previous optical studies. 
We found that they are above the galaxy main sequence, suggesting a starburst nature. 
This suggests a higher fraction of starburst galaxies in SLSN hosts than estimated in previous studies. 
We calculated the extinction at H$\alpha$ wavelength from the ratio between radio SFRs and dust-uncorrected H$\alpha$ SFRs. 
The SLSN hosts are on the trend of increasing extinction with metallicity, which is seen in local star-forming galaxies. 
The PTF~10qaf host has a higher extinction compared to the sample of star-forming galaxies, suggesting that it has highly obscured star formation. 
Because the sample in this study is still limited, it is important to increase the sample size covering a wider range of SFR and stellar mass in order to understand the general properties of SLSN hosts,

Our radio observations also place a constraint on a pulsar-driven SN model which predicts quasi-steady radio emission. 
We found that our radio 3$\sigma$ upper limit on the SN~2010gx host is inconsistent with the model of \cite{oman18} in the case of no absorption processes by assuming a synchrotron self-absorption spectral index of 2.5. 
Because the radio emission is predicted to reach its peak at around 10 years after the explosion, long-term follow-up observations are important to constrain the model with different parameters.

\acknowledgments

We would like to acknowledge NRAO staffs for their help in preparation of observations. 
We thank the referee for helpful comments and suggestions. 
We are grateful to the PDJ collaboration for providing opportunities for fruitful discussions. 
BH and TM are supported by JSPS KAKENHI Grant Number 15K17616, 16H02158, and 25800103. 
The National Radio Astronomy Observatory is a facility of the National Science Foundation operated under cooperative agreement by Associated Universities, Inc.
Based on observations made with the NASA/ESA Hubble Space Telescope, obtained from the Data Archive at the Space Telescope Science Institute, which is operated by the Association of Universities for Research in Astronomy, Inc., under NASA contract NAS 5-26555. 
Funding for SDSS-III has been provided by the Alfred P. Sloan Foundation, the Participating Institutions, the National Science Foundation, and the U.S. Department of Energy Office of Science. The SDSS-III web site is http://www.sdss3.org/.
SDSS-III is managed by the Astrophysical Research Consortium for the Participating Institutions of the SDSS-III Collaboration including the University of Arizona, the Brazilian Participation Group, Brookhaven National Laboratory, Carnegie Mellon University, University of Florida, the French Participation Group, the German Participation Group, Harvard University, the Instituto de Astrofisica de Canarias, the Michigan State/Notre Dame/JINA Participation Group, Johns Hopkins University, Lawrence Berkeley National Laboratory, Max Planck Institute for Astrophysics, Max Planck Institute for Extraterrestrial Physics, New Mexico State University, New York University, Ohio State University, Pennsylvania State University, University of Portsmouth, Princeton University, the Spanish Participation Group, University of Tokyo, University of Utah, Vanderbilt University, University of Virginia, University of Washington, and Yale University.

\facility{VLA}



\begin{thebibliography}{}
\bibitem[Angus et al.(2016)]{angu16} Angus, C.~R., Levan, A.~J., Perley, D.~A., et al.\ 2016, \mnras, 458, 84 
\bibitem[Becker et al.(1995)]{beck95} Becker, R.~H., White, R.~L., \& Helfand, D.~J.\ 1995, \apj, 450, 559 
\bibitem[Bock et al.(1999)]{bock99} Bock, D.~C.-J., Large, M.~I., \& Sadler, E.~M.\ 1999, \aj, 117, 1578 
\bibitem[Calzetti et al.(2000)]{calz00} Calzetti, D., Armus, L., Bohlin, R.~C., et al.\ 2000, \apj, 533, 682 
\bibitem[Chabrier(2003)]{chab03} Chabrier, G.\ 2003, \pasp, 115, 763 
\bibitem[Chevalier \& Irwin(2011)]{chev11} Chevalier, R.~A., \& Irwin, C.~M.\ 2011, \apjl, 729, L6 
\bibitem[Cikota et al.(2017)]{ciko17} Cikota, A., De Cia, A., Schulze, S., et al.\ 2017, \mnras, 469, 4705 
\bibitem[Condon et al.(1998)]{cond98} Condon, J.~J., Cotton, W.~D., Greisen, E.~W., et al.\ 1998, \aj, 115, 1693 
\bibitem[Condon(1992)]{cond92} Condon, J.~J.\ 1992, \araa, 30, 575 
\bibitem[Cooke et al.(2012)]{cook12} Cooke, J., Sullivan, M., Gal-Yam, A., et al.\ 2012, \nat, 491, 228 
\bibitem[Dexter \& Kasen(2013)]{dext13} Dexter, J., \& Kasen, D.\ 2013, \apj, 772, 30 
\bibitem[Duric et al.(1988)]{duri88} Duric, N., Bourneuf, E., \& Gregory, P.~C.\ 1988, \aj, 96, 81 
\bibitem[Elbaz et al.(2011)]{elba11} Elbaz, D., Dickinson, M., Hwang, H.~S., et al.\ 2011, \aap, 533, A119 
\bibitem[Gal-Yam(2012)]{gal12} Gal-Yam, A.\ 2012, Science, 337, 927 
\bibitem[Gal-Yam et al.(2009)]{gal09} Gal-Yam, A., Mazzali, P., Ofek, E.~O., et al.\ 2009, \nat, 462, 624 
\bibitem[Garn \& Best(2010)]{garn10} Garn, T., \& Best, P.~N.\ 2010, \mnras, 409, 421 
\bibitem[Gioia et al.(1982)]{gioi82} Gioia, I.~M., Gregorini, L., \& Klein, U.\ 1982, \aap, 116, 164 
\bibitem[Greiner et al.(2016)]{grei16} Greiner, J., Micha{\l}owski, M.~J., Klose, S., et al.\ 2016, \aap, 593, A17 
\bibitem[Hancock et al.(2012)]{hanc12} Hancock, P.~J., Murphy, T., Gaensler, B.~M., Hopkins, A., \& Curran, J.~R.\ 2012, \mnras, 422, 1812
\bibitem[Hatsukade et al.(2012)]{hats12} Hatsukade, B., Hashimoto, T., Ohta, K., et al.\ 2012, \apj, 748, 108 
\bibitem[Hjorth et al.(2003)]{hjor03} Hjorth, J., et al.\ 2003, \nat, 423, 847 
\bibitem[Japelj et al.(2016)]{jape16} Japelj, J., Vergani, S.~D., Salvaterra, R., Hunt, L.~K., \& Mannucci, F.\ 2016, \aap, 593, A115 
\bibitem[Kasen \& Bildsten(2010)]{kase10} Kasen, D., \& Bildsten, L.\ 2010, \apj, 717, 245 
\bibitem[Kashiyama \& Murase(2017)]{kash17} Kashiyama, K., \& Murase, K.\ 2017, \apjl, 839, L3 
\bibitem[Kennicutt(1998)]{kenn98} Kennicutt, R.~C., Jr.\ 1998, \apj, 498, 541
\bibitem[Klein et al.(2017)]{klei17} Klein, U., Lisenfeld, U., \& Verley, S.\ 2017, arXiv:1710.03149 
\bibitem[Kroupa(2001)]{krou01} Kroupa, P.\ 2001, \mnras, 322, 231 
\bibitem[Leloudas et al.(2015)]{lelo15} Leloudas, G., Schulze, S., Kr{\"u}hler, T., et al.\ 2015, \mnras, 449, 917 
\bibitem[Lunnan et al.(2014)]{lunn14} Lunnan, R., Chornock, R., Berger, E., et al.\ 2014, \apj, 787, 138 
\bibitem[Madau \& Dickinson(2014)]{mada14} Madau, P., \& Dickinson, M.\ 2014, \araa, 52, 415 
\bibitem[McMullin et al.(2007)]{mcmu07} McMullin, J.~P., Waters, B., Schiebel, D., Young, W., \& Golap, K.\ 2007, Astronomical Data Analysis Software and Systems XVI, 376, 127 
\bibitem[Metzger et al.(2017)]{metz17} Metzger, B.~D., Berger, E., \& Margalit, B.\ 2017, \apj, 841, 14 
\bibitem[Micha{\l}owski et al.(2014)]{mich14} Micha{\l}owski, M.~J., Hayward, C.~C., Dunlop, J.~S., et al.\ 2014, \aap, 571, A75 
\bibitem[Micha{\l}owski et al.(2012)]{mich12} Micha{\l}owski, M.~J., Kamble, A., Hjorth, J., et al.\ 2012, \apj, 755, 85 
\bibitem[Micha{\l}owski et al.(2015)]{mich15} Micha{\l}owski, M.~J., Gentile, G., Hjorth, J., et al.\ 2015, \aap, 582, A78 
\bibitem[Moriya et al.(2010)]{mori10} Moriya, T., Tominaga, N., Tanaka, M., Maeda, K., \& Nomoto, K.\ 2010, \apjl, 717, L83
\bibitem[Moriya et al.(2013)]{mori13} Moriya, T.~J., Blinnikov, S.~I., Tominaga, N., et al.\ 2013, \mnras, 428, 1020 
\bibitem[Murase et al.(2016)]{mura16} Murase, K., Kashiyama, K., \& M{\'e}sz{\'a}ros, P.\ 2016, \mnras, 461, 1498 
\bibitem[Murphy et al.(2011)]{murp11} Murphy, E.~J., Condon, J.~J., Schinnerer, E., et al.\ 2011, \apj, 737, 67 
\bibitem[Niklas et al.(1997)]{nikl97} Niklas, S., Klein, U., \& Wielebinski, R.\ 1997, \aap, 322, 19 
\bibitem[Omand et al.(2018)]{oman18} Omand, C.~M.~B., Kashiyama, K., \& Murase, K.\ 2018, \mnras, 474, 573 
\bibitem[Perley \& Perley(2013)]{pp13} Perley, D.~A., \& Perley, R.~A.\ 2013, \apj, 778, 172 
\bibitem[Perley et al.(2013)]{perl13} Perley, D.~A., Levan, A.~J., Tanvir, N.~R., et al.\ 2013, \apj, 778, 128 
\bibitem[Perley et al.(2015)]{perl15} Perley, D.~A., Perley, R.~A., Hjorth, J., et al.\ 2015, \apj, 801, 102 
\bibitem[Perley et al.(2016)]{perl16} Perley, D.~A., Quimby, R.~M., Yan, L., et al.\ 2016, \apj, 830, 13 
\bibitem[Perley \& Butler(2017)]{pb17} Perley, R.~A., \& Butler, B.~J.\ 2017, \apjs, 230, 7 
\bibitem[Planck Collaboration et al.(2016)]{plan16} Planck Collaboration, Ade, P.~A.~R., Aghanim, N., et al.\ 2016, \aap, 594, A13 
\bibitem[Perley et al.(2017)]{perl17} Perley, D.~A., Hjorth, J., Tanvir, N.~R., \& Perley, R.~A.\ 2017, \mnras, 465, 970 
\bibitem[Pettini \& Pagel(2004)]{pett04} Pettini, M., \& Pagel, B.~E.~J.\ 2004, \mnras, 348, L59 
\bibitem[Rodighiero et al.(2011)]{rodi11} Rodighiero, G., Daddi, E., Baronchelli, I., et al.\ 2011, \apjl, 739, L40 
\bibitem[Salpeter(1955)]{salp55} Salpeter, E.~E.\ 1955, \apj, 121, 161 
\bibitem[Schulze et al.(2018)]{schu18} Schulze, S., Kr{\"u}hler, T., Leloudas, G., et al.\ 2018, \mnras, 473, 1258 
\bibitem[Smol{\v c}i{\'c} et al.(2017a)]{smol17a} Smol{\v c}i{\'c}, V., Delvecchio, I., Zamorani, G., et al.\ 2017a, \aap, 602, A2 
\bibitem[Smol{\v c}i{\'c} et al.(2017b)]{smol17b} Smol{\v c}i{\'c}, V., Novak, M., Bondi, M., et al.\ 2017b, \aap, 602, A1 
\bibitem[Sorokina et al.(2016)]{soro16} Sorokina, E., Blinnikov, S., Nomoto, K., Quimby, R., \& Tolstov, A.\ 2016, \apj, 829, 17 
\bibitem[Speagle et al.(2014)]{spea14} Speagle, J.~S., Steinhardt, C.~L., Capak, P.~L., \& Silverman, J.~D.\ 2014, \apjs, 214, 15 
\bibitem[Stanek et al.(2003)]{stan03} Stanek, K.~Z., et al.\ 2003, \apjl, 591, L17 
\bibitem[Stanway et al.(2010)]{stan10} Stanway, E.~R., Davies, L.~J.~M., \& Levan, A.~J.\ 2010, \mnras, 409, L74 
\bibitem[Stanway et al.(2014)]{stan14} Stanway, E.~R., Levan, A.~J., \& Davies, L.~J.~M.\ 2014, \mnras, 444, 2133 
\bibitem[Stanway et al.(2015)]{stan15} Stanway, E.~R., Levan, A.~J., Tanvir, N., et al.\ 2015, \mnras, 446, 3911 
\bibitem[Tabatabaei et al.(2017)]{taba17} Tabatabaei, F.~S., Schinnerer, E., Krause, M., et al.\ 2017, \apj, 836, 185
\bibitem[Woosley(2010)]{woos10} Woosley, S.~E.\ 2010, \apjl, 719, L204 
\bibitem[Woosley et al.(2007)]{woos07} Woosley, S.~E., Blinnikov, S., \& Heger, A.\ 2007, \nat, 450, 390 
\end{thebibliography}
\end{document}